\documentclass[conference]{IEEEtran}
\IEEEoverridecommandlockouts
\usepackage{amsmath,amssymb,amsfonts}
\usepackage{graphicx}
\usepackage{textcomp}
\usepackage{xcolor}

\usepackage[style=ieee, url=false]{biblatex}
\AtBeginBibliography{\small}

\usepackage{xcolor}
\usepackage{centernot}
\usepackage{braket}
\usepackage{caption}
\usepackage{subcaption}
\usepackage{hyperref}

\usepackage{algorithm}
\usepackage{algpseudocode}

\algnewcommand{\algorithmicand}{\textbf{ and }}
\algnewcommand{\algorithmicor}{\textbf{ or }}
\algnewcommand{\OR}{\algorithmicor}
\algnewcommand{\AND}{\algorithmicand}

\renewcommand\vec{\mathbf}

\DeclareMathOperator{\rank}{rank}

\begin{document}

\title{Online Gaussian elimination for quantum LDPC decoding}

\author{
\IEEEauthorblockN{Sam J. Griffiths, Asmae Benhemou, Dan E. Browne}
\IEEEauthorblockA{
\textit{Department of Physics \& Astronomy, University College London, London, WC1E 6BT, United Kingdom}}
}

\maketitle

\begin{abstract}
Decoders for quantum LDPC codes generally rely on solving a parity-check equation with Gaussian elimination, with the generalised union--find decoder performing this repeatedly on growing clusters. We present an online variant of the Gaussian elimination algorithm which maintains an LUP decomposition in order to process only new rows and columns as they are added to a system of equations. This is equivalent to performing Gaussian elimination once on the final system of equations, in contrast to the multiple rounds of Gaussian elimination employed by the generalised union--find decoder. It thus significantly reduces the number of operations performed by the decoder. We consider the generalised union--find decoder as an example use case and present a complexity analysis demonstrating that both variants take time cubic in the number of qubits in the general case, but that the number of operations performed by the online variant is lower by an amount which itself scales cubically. This analysis is also extended to the regime of `well-behaved' codes in which the number of growth iterations required is bounded logarithmically in error weight. Finally, we show empirically that our online variant outperforms the original offline decoder in average-case time complexity on codes with sparser parity-check matrices or greater covering radius.
\end{abstract}

\maketitle


\section{Introduction} \label{sec:introduction}
Gaussian elimination is a well-known algorithm for taking a matrix into row echelon form, which can be used as a general method for solving systems of linear equations \cite{shafarevichLinearAlgebraGeometry2012}. These systems are ubiquitous across many problem domains -- one such example is in error-correcting codes and, in particular, quantum error correction.

Active quantum error correction is expected to carry an essential role in the fault-tolerant storage and processing of quantum information~\cite{dennis2002topological, nielsenQuantumComputationQuantum2002, gottesmanIntroductionQuantumError2002}. This consists of reliably encoding logical information in the joint Hilbert space of a collection of physical systems \cite{shor1996fault, preskill1998fault, dennis2002topological}. When information is corrupted under a noise channel acting on these qubits, a classical decoding algorithm is used to interpret collected parity measurements over the physical qubits, to infer a recovery operation \cite{dennis2002topological, delfosse2021almost}. Designing high-performance and practical decoding algorithms is crucial for reducing the space and time resource costs of logical information encoding, enabling the achievement of a target logical error rate with fewer physical qubits~\cite{roffeDecodingQuantumLDPC2020, skoric2023parallel, ai2024quantum}.

Quantum low-density parity-check (qLDPC) codes encompass families of quantum code constructions with bounded stabilizer weight, high information storage capacity, and distance scaling logarithmically or faster with the physical system size \cite{tillich2013quantum, panteleevDegenerateQuantumLDPC2021, panteleev2022asymptotically}. The practical use of classical LDPC codes can be attributed to the existence of exceptionally fast decoding algorithms; in particular, belief propagation (BP) offers a linear-time complexity with performance nearing that of the maximum-likelihood decoder \cite{mceliece1998turbo, mackay1997near, richardson2001capacity}. In contrast, the structure of quantum LDPC codes (specifically, the existence of more than one optimal solution) leads such approaches to be ineffective without modification \cite{roffeDecodingQuantumLDPC2020}.

In its most foundational form, the decoding problem is one of solving a linear system of parity-check equations of the form $H\vec{x}=\vec{\sigma}$ \cite{macwilliamsTheoryErrorcorrectingCodes1977}. Decoders for qLDPC codes frequently rely on Gaussian elimination to directly solve this equation. Approaches to optimising the solution found include iteratively growing clusters from a minimum solution neighbourhood, as in the generalised union--find decoder \cite{delfosse2022toward}, and using belief propagation to inform the choice of free variables, as in BP ordered-statistics decoding (BP-OSD) \cite{fossorierSoftDecisionDecoding1997, panteleevDegenerateQuantumLDPC2021, roffeDecodingQuantumLDPC2020}.

In the generalised union--find decoder, Gaussian elimination is performed repeatedly on a strictly-increasing subset of the global linear system. In this work, we present an online variant of the Gaussian elimination algorithm, which dynamically maintains a row echelon form whilst new rows and columns are added to the system. This is equivalent to performing Gaussian elimination once on the final system and thus has the potential to significantly reduce the overall number of operations performed. In Section~\ref{sec:background} we review the relevant background of Gaussian elimination, quantum codes and decoders. In Section~\ref{sec:results} we present the online Gaussian elimination procedure and, as a case study, outline the time complexity of the generalised union--find decoder using this technique, alongside empirical results on three different quantum error-correcting codes. Finally, we offer a discussion and concluding remarks in Section~\ref{sec:conclusion}.

In this work, we list pseudocode for outlining Gaussian elimination and our online variant. Algorithm~\ref{alg:lup} describes LUP decomposition as exactly the same procedure as Gaussian elimination but with extra logging. Algorithm~\ref{alg:online-lup} describes an online algorithm for adding new rows and columns to an existing LUP decomposition.

\section{Background} \label{sec:background}
\subsection{Gaussian elimination} \label{subsec:gaussian}
Consider a system of linear equations of the form
\begin{equation} \label{eq:linear-system}
    A\vec{x} = \vec{b} \ ,
\end{equation}
where $A$ is the $m \times n$ coefficient matrix representing $m$ equations and $n$ variables. Gaussian elimination is an algorithm which takes a matrix into row echelon form (REF), which is defined as a matrix in which the first nonzero entry of each row, known as the \emph{pivot}, is to the right of the pivots of all rows above. Rows containing only zeroes are relegated to the bottom of the matrix. The algorithm takes a matrix into REF by repeatedly applying three elementary row operations:
\begin{enumerate}
    \item Swapping two rows;
    \item Multiplying a row by a constant;
    \item Subtracting a multiple of a row from another row.
\end{enumerate}
The $\mathbb{Z}_2$ field (i.e.\ binary with addition/subtraction modulo 2) is arguably the simplest possible variant of Gaussian elimination. Subtraction modulo 2 is merely the XOR operation, and the only possible nonzero constant is 1, simplifying Operation 3 and removing Operation 2 entirely. We will later see how the $\mathbb{Z}_2$ field is relevant for the case of error-correcting codes.

The row echelon form reveals useful information about the matrix and its dimensionality. For example, the \emph{rank} of a matrix, defined as the number of linearly independent rows (or, equivalently, columns) is trivially the number of nonzero rows in the REF. To solve a system of the form in Equation~\ref{eq:linear-system}, we first define the \emph{augmented} matrix $A|\vec{b}$ as the matrix $A$ with the vector $\vec{b}$ appended as an additional column. As per the Rouché--Capelli theorem, the system is \emph{inconsistent}, i.e.\ has no solutions, if $\rank(A|\vec{b}) > \rank(A)$, but it is \emph{consistent}, i.e.\ has one or more solutions, if $\rank(A|\vec{b}) = \rank(A)$ \cite{shafarevichLinearAlgebraGeometry2012}. Both of these ranks can be calculated by performing Gaussian elimination on $A|\vec{b}$. If the system is consistent, then a solution(s) can be obtained by \emph{back-substitution}: the lowermost equation in REF contains only one unknown and so is trivially solvable, which is then substituted into the equation above such that it too contains only one unknown, and so on. If $\rank(A|\vec{b})=\rank(A)=r$ and $n=r$, then the system is \emph{determined} and there exists a unique solution, but if $n>r$, then it is \emph{underdetermined} and there exist infinitely many solutions generated by $n-r$ free variables (or, in the case of $\mathbb{Z}_2$, a finite yet exponentially large number of solutions).

\subsection{Linear codes}
Classically, an $[n,k,d]$ code encodes $k$ bits' worth of logical information into $n$ physical bits, where $n>k$ in order to introduce redundancy to protect the bulk from errors (i.e.\ bit-flips), and the \emph{code distance} $d$ is defined as the minimum Hamming distance between two codewords, or equivalently the minimum number of physical errors needed to form a logical error \cite{macwilliamsTheoryErrorcorrectingCodes1977}. For example, the $[3,1,3]$ repetition code encodes a single logical bit as
\begin{equation}
    0_L = 000 \ , \ 1_L = 111 \ .
\end{equation}

When a received word is outside of the codespace $\{000,111\}$, one or more errors must have occurred, and a majority vote can be performed. Given a prior error rate $p$, i.e.\ the independent chance of a bit-flip on each physical bit, this reduces the overall logical error rate $p_L$ to
\begin{equation}
    p_L = p^3 + 3p^2(1-p) \ .
\end{equation}
For higher $n$, $p_L$ follows a binomial expression which is suppressed to zero as $n\to\infty$ for $p$ below a threshold.

In the general case, an $[n,k,d]$ code is said to be \emph{linear} if every linear combination of codewords is itself a codeword. Logical bitstrings $\vec{u}=u_1u_2 \ldots u_k$ are encoded into codewords $\vec{x}=x_1x_2 \ldots x_k$ by a generator matrix $G$ as
\begin{equation}
    \vec{u}G = \vec{x} \ .
\end{equation}

To detect errors, one or more parity checks (i.e.\ sum modulo 2) are performed on the physical bits. This is represented by a parity-check matrix $H$, where each row corresponds to a parity check on a subset of the physical bits. These parity checks yield a result such that
\begin{equation} \label{eq:parity-check}
    H\vec{x}^\top = \vec{\sigma} \ ,
\end{equation}
where $\vec{\sigma}$ is the \emph{syndrome} of the error, sometimes denoted $\vec{\sigma}(\vec{x})$. A word $\vec{x}$ is in the codespace if and only if $\vec{\sigma}(\vec{x}) = H\vec{x}^\top = \vec{0}$.

To correct errors, we must correctly predict the error state $\vec{x}$ given the syndrome $\vec{\sigma}$. This is fundamentally equivalent to solving for $\vec{x}$ in Equation~\ref{eq:parity-check}, which is a linear system of the form in Equation~\ref{eq:linear-system}. Therefore, Gaussian elimination can be used to solve the decoding problem in general. However, it does not suffice to find any arbitrary solution: we wish to find the most likely error consistent with the syndrome. Decoding algorithms therefore attempt to find or approximate optimal solutions and may or may not utilise Gaussian elimination to this end \cite{fossorierSoftDecisionDecoding1997, macwilliamsTheoryErrorcorrectingCodes1977, delfosse2022toward, panteleevDegenerateQuantumLDPC2021, roffeDecodingQuantumLDPC2020}.

\subsection{Quantum codes}

A quantum code with parameters $[[n,k,d]]$ defined on a set of $n$ physical qubits is a linear subspace of a Hilbert space of dimension $2^n$ encoding $k$ logical qubits, with code distance $d$. In particular, a \emph{Calderbank--Shor--Steane} (CSS) code~\cite{calderbank1996good, shor1996fault} is generally defined using a pair of classical linear codes $C_X, C_Z \subseteq \mathbb{F}_2^n$ satisfying the orthogonality condition $C_Z^{\perp} \subseteq C_X$. Such a code has length dependent on the lengths of $C_X$ and $C_Z$, a logical subspace encoding $k = \text{dim} \left( C_X \backslash C_Z^{\perp} \right)$ qubits, and distance $d = \text{min}(d_X,d_Z)$, where $d_X$ and $d_Z$ are the minimum Hamming weights of vectors from $C_X \backslash C_Z^{\perp}$ and $C_Z \backslash C_X^{\perp}$ respectively. A code $\mathcal{Q}$ can be represented using a parity-check matrix given by the check matrices of the binary codes $C_X$ and $C_Z$, i.e.\ $H \equiv (H_X, H_Z)$, where the rows of $H$ define the generators of the stabilizer group of the quantum code. The orthogonality condition can then be expressed as $H_XH_Z^T=0$. Of particular interest are stabilizer codes defined using sparse parity-check matrices, i.e.\ quantum low-density parity-check (qLDPC) codes, whose structure allows for fewer operations to detect and correct errors. Topological codes are another significant class of quantum codes, which encode logical qubits in the global degrees of freedom of a physical system~\cite{dennis2002topological, kitaevFaulttolerantQuantumComputation2003, terhal2015quantum, bombin2006topological}.

In this work, we evaluate decoder implementations on three topological codes satisfying the qLDPC property. First, the 2D toric code is a CSS code defined on a two-dimensional square lattice, with qubits placed on the edges of the lattice with periodic boundary conditions~\cite{kitaevFaulttolerantQuantumComputation2003}. The $X$ and $Z$ stabilizer checks are respectively generated by four-body measurement operators on the qubits located on the edges around each vertex $v$ and face $f$ of the lattice, namely
\begin{equation}
    S_v^X = \prod_{e \in \partial v} X_e \ , \ S_f^Z = \prod_{e \in \partial f} Z_e \ .
\end{equation}
The toric code has parameters $[[2L^2, 2, L]]$, where the logical operators correspond to non-trivial loops wrapping around the torus and $L$ is the side length of the square lattice. This construction generalises to cellulations of $D$-dimensional tori; in particular, the 3D toric code is defined by placing the qubits on the edges of a cubic lattice, and establishing the $X$ and $Z$ stabiliser checks in the same manner as the 2D toric code above. Using periodic boundary conditions, the code defined by these operators has parameters $[[3L^3, 3, d_X=L^2, d_Z=L]]$ where $L$ is the side length of the cubic lattice~\cite{castelnovo2008topological}. Finally, the 2D colour code is defined via the cellulation of a three-colourable two-manifold, originally defined on a hexagonal $(6.6.6)$ lattice~\cite{bombin2006topological}. The qubits are placed on the vertices of the lattice, and the stabilizer group defining the code is generated by the face operators 
\begin{equation}
    S_f^X = \prod_{v \in \partial f} X_v \ , \ S_f^Z = \prod_{v \in \partial f} Z_v \ ,
\end{equation}
for every hexagonal face $f$, where $v \in \partial f$ represents the qubits on the vertices around a face. On a triangular geometry with open boundary conditions, the colour code has parameters $[[3(d^2 - 1)/4, 1, d]]$, but we will hereinafter consider periodic boundary conditions for simplicity and consistency with the above codes.

In quantum error correction, errors affecting a collection of qubits are typically modelled through Pauli channels. Similar to classical error correction, the stabilizer checks of a quantum code are measured to detect such errors, and the resulting measurements form the syndrome $\vec
{\sigma}$, a classical bit string that indicates which stabilizer checks have been violated. Given an error $\vec{e}$, solving the decoding problem consists of identifying an estimated error $\vec{e}'$, which gives rise to the same syndrome $\sigma$, while minimising weight support. In the case of CSS codes, error detection and correction is based on the sets of stabilizer checks generated by the rows of $H_X$ and $H_Z$, which respectively detect phase-flip and bit-flip errors.

\subsection{Union--find decoder}
The union--find (UF) decoder was introduced in \cite{delfosse2021almost} and \cite{delfosse2020linear} as a near-linear-time decoder for surface codes \cite{kitaevFaulttolerantQuantumComputation2003}, relying on parity conditions and a peeling algorithm for efficient error correction. Succinctly, clusters are grown uniformly across the lattice from each element in the syndrome $\sigma$ until they support an even number of syndrome elements, at which point they are reduced (`peeled') down to a correction operator via simple state machine rules. It has since been shown that the UF decoder runs in strictly linear time under independent and phenomenological noise models even without the algorithmic optimisations implied by its namesake \cite{griffithsUnionfindQuantumDecoding2024}. Graphically, the UF decoder is equivalent to approximating a minimum-weight perfect matching by growing clusters within the neighbourhood of the error syndrome \cite{griffithsUnionfindQuantumDecoding2024}.

UF has since been extended beyond surface codes to more general quantum LDPC code constructions \cite{delfosse2022toward}. The original UF decoder performs well on surface codes due to the strictly graphlike nature of their decoding graphs; however, the decoding graphs on general LDPC codes are instead hypergraphs. Notions of cluster parity and state-machine peeling do not easily generalise to the case of hypergraphs: a cluster no longer necessarily contains a valid solution simply by having even parity, and the peeling process generalises to a costly combinatorial search. Therefore, it appears that Gaussian elimination must be relied upon for two black-box subroutines used by the generalised decoder: \emph{syndrome validation} and \emph{solution generation}. These subproblems respectively constitute using Equation~\ref{eq:parity-check} to check for the existence of a solution, and finding a solution. By growing clusters from $\vec{\sigma}$, Gaussian elimination is performed repeatedly on a strictly-increasing subset of the global linear system; that is, rows and columns are iteratively added to the augmented matrix $H|\vec{\sigma}$ and Gaussian elimination is performed anew each time. In the following section, we propose a method for an online approach to Gaussian elimination which only needs to process new rows and columns as they are added.

We refer readers to \cite{delfosse2022toward} for a more rigorous introduction to the generalised UF decoder. Importantly, the authors of \cite{delfosse2022toward} showed that generalised UF performs well for certain classes of qLDPC codes, which we will call `well-behaved' codes. Notably, these were shown to include topological codes, namely higher-dimensional ($D \geq 3$) toric and hyperbolic codes, and locally-testable codes. This analysis introduced a quantity termed the \emph{covering radius} $\rho_\text{cov}(\vec{\sigma})$ of a syndrome $\vec{\sigma}$, which is defined as the number of growth steps needed by the algorithm to cover a valid solution to $\sigma$. We build on their argument in the complexity analysis in Section~\ref{subsec:complexity}. 

\section{Results}
\label{sec:results}
\subsection{Online Gaussian elimination} \label{subsec:online-gaussian}
Let us take the generalised union--find decoder as a case study. First, we define $H'$ as the `reduced' parity-check matrix $H$, filtered to contain only rows and columns representing the checks and variables, respectively, currently contained in the interior (i.e.\ non-boundary) of any cluster. Gaussian elimination solves the problem in the general case: one or more solutions exist, and thus clusters can stop growing, when $\rank(H'|\vec{\sigma}) = \rank(H') = r$, yielding a solution generator with $n-r$ free variables, as we can expect degeneracy (i.e.\ underdetermination) in the general case.

The decoder is initialised with the syndrome $\vec{\sigma}$ and neighbouring nodes are iteratively added until a solution(s) exists. By definition, $\rho_\text{cov}(\vec{\sigma})$ growth steps are required, meaning that Gaussian elimination is performed this many times on a matrix $H'|\vec{\sigma}$ of strictly increasing size.

Instead, we propose an online variant of Gaussian elimination which removes redundant computational work between growth steps. An \emph{online algorithm} is one in which a valid solution is maintained as new input is obtained over time, i.e.\ the final problem data is not required in whole to commence work \cite{borodinOnlineComputationCompetitive2005}.

Let $H_0$ be an augmented matrix in REF from a previous growth step and $H_1$ be the same matrix with additional rows and columns appended. The new data must be brought up-to-date with decisions made in previous growth steps. We record decisions made by the rounds of Gaussian elimination by maintaining an LUP decomposition. This is equivalent to Gaussian elimination, except the elementary row operations are explicitly represented by a matrix factorisation of the form
\begin{equation}
    PA = LU \ ,
\end{equation}
where $A$ is the original matrix and $U$ is the matrix in row echelon form (upper-triangular). Swapped rows are recorded by the permutation matrix $P$ and row subtractions are recorded by the lower-triangular matrix $L$. Algorithm~\ref{alg:lup} shows how this factorisation is equivalent to performing Gaussian elimination whilst recording decisions.

Firstly, by maintaining the matrix factors $P$ and $L$, previous row operations can be performed on new rows and columns when they are added to the system. Secondly, by definition of the decoder, $H_0$ represents an inconsistent system, suggesting the existence of `missing' pivots (i.e.\ zeroes) from the leading diagonal of $U$. These positions are candidates for pivots to be found within the newly-added rows. Commencing from the first missing pivot position, Gaussian elimination is performed, except that only the newly-added rows need to be searched through and, by extension, subtracted from. This is justified as $H_0$ is already in REF and is thus upper-triangular, such that only zeroes can be present beneath the leading diagonal in the old rows. In this way, the LUP decomposition is updated in each growth step to return $U$ into REF given the new rows and columns. Algorithm~\ref{alg:online-lup} lists pseudocode for this online LUP decomposition update performed in each growth step.

\subsection{Complexity analysis} \label{subsec:complexity}
For a square  $n \times n$ matrix (as arises with an exactly-determined system of equations) the time complexity of Gaussian elimination is $O(n^3)$. More generally, for an $m \times n$ (i.e.\ rectangular) matrix, the time complexity is $O(mn \min(m,n))$, a.k.a.\ \emph{big-times-small-squared} \cite{boydIntroductionAppliedLinear2018}. It is straightforward to see how the cubic complexity arises. For each of the $n$ pivots, the row is subtracted from $O(n)$ other rows, which each contain $n$ elements. No order of complexity is added by obtaining an LUP decomposition, as this amounts to merely logging the operations which have been performed (Algorithm~\ref{alg:lup}). Finally, back-substitution is trivially $O(n^2)$. 

In contrast, the online LUP update (Algorithm~\ref{alg:online-lup}), for each of the $O(n)$ outstanding pivot positions, needs only search through and subtract from the newly-added rows. Therefore, it is asymptotically equivalent to performing a single LUP decomposition on the final-sized system.

In this section, we abide by the notation and framework introduced for the generalised UF decoder in \cite{delfosse2022toward}. The union of all clusters of parity checks and qubits (i.e.\ vertices and hyperedges) is denoted the \emph{erasure} $\mathcal{E}$ (this terminology is a result of the decoder's earliest description on the erasure noise channel \cite{delfosse2020linear, delfosse2021almost}). At the start of the algorithm, this is initialised as $\mathcal{E}=\sigma$.

Once one or more solutions exist, the clusters stop growing and the final reduced parity-check matrix $H'$ has dimensions $r \times c$ where $r+c=|\mathcal{E}|$. The number of XOR operations required by Gaussian elimination on $H'$ -- and thus by the online decoder -- is $O(|\mathcal{E}|^3)$. In the worst case, $|\mathcal{E}|=n$, taking $n$ to be the total of both qubits and checks in the code to simplify analysis. This gives our online decoder a worst-case complexity of $O(n^3)$.

This can be contrasted with the original offline description of the decoder in \cite{delfosse2022toward}. Gaussian elimination is performed anew for each of the $\rho_\text{cov}(\vec{\sigma})$ growth steps (denoted $\rho$ for concision), in each of which the size of the erasure is increased by $O(\delta)$, where $\delta$ is the maximum degree of the Tanner graph. The worst-case number of operations is now approximately given by
\begin{equation}
    \sum_{i=0}^{\rho} \left( \frac{n}{\delta^i} \right)^3 = \frac{n^3 \delta^{-3\rho} (\delta^{3\rho+3}-1)}{\delta^3-1} \ ,
\end{equation}
from which it follows that the offline decoder also has a worst-case complexity of $O(n^3)$. However, it is apparent that the online decoder has significantly reduced overhead; indeed, the number of operations skipped by the online variant is given by
\begin{equation}
     \sum_{i=0}^{\rho-1} \left( \frac{n}{\delta^i} \right)^3 = \frac{n^3 \delta^{3-3\rho} (\delta^{3\rho}-1)}{\delta^3-1} \ ,
\end{equation}
which is itself $O(n^3)$.

This analysis can be refined in the case of codes with a property identified in \cite{delfosse2022toward} which we call `well-behaved' codes. These are codes where $\rho \leq C\log|\vec{x}|$ for all $|\vec{x}|<w$ for some constants $C,w$. In this case, the decoder corrects all errors where $|\vec{x}| < \min(w, Ad^\alpha)$, where $A$ and $\alpha$ are constants which depend on the degree of the Tanner graph (see Equation~\ref{eq:a-d-alpha}), $d$ is the code distance, and the erasure formed is bounded by
\begin{align}
    |\mathcal{E}| &\leq \delta^2|\vec{x}| \cdot \delta^{\rho} \\
    &\leq \delta^2 |\vec{x}|^{1+C\log\delta} \ .
\end{align}
The number of operations performed by the online decoder is approximately
\begin{align}
    |\mathcal{E}|^3 \leq \delta^6 |\vec{x}|^{3+3C\log\delta} \ ,
\end{align}
and thus the number performed by the offline decoder is approximately
\begin{align}
    (\delta^6|\vec{x}|^3)(1 + |\vec{x}|^3 + \cdots + |\vec{x}|^{3C\log\delta}) \ ,
\end{align}
where the upper bound for $|\vec{x}|$ varies with code-dependent properties. In this well-behaved regime, it is bounded by $Ad^\alpha$, defined as
\begin{equation} \label{eq:a-d-alpha}
    Ad^\alpha = \left(\frac{d}{2\delta^2}\right)^\frac{1}{1+C\log\delta} \ .
\end{equation}
This suggests an upper bound for the size of the erasure formed as
\begin{equation}
    |\mathcal{E}| \leq \delta^2 \cdot \frac{d}{2\delta^2} = \frac{d}{2} \ ,
\end{equation}
which implies that the online decoder has a complexity of $O(d^3)$ and that the number of operations skipped in contrast with the offline decoder is approximately
\begin{align}
    (\delta^2 |\vec{x}|^{1+C\log\delta-1})^3 &= \left(\frac{d}{2|\vec{x}|}\right)^3 \\
    &= \left(\frac{d}{2Ad^\alpha}\right)^3 \ .
\end{align}

\subsection{Simulation}
\begin{figure}
    \centering

    \begin{subfigure}{\linewidth}
        \includegraphics[width=\linewidth]{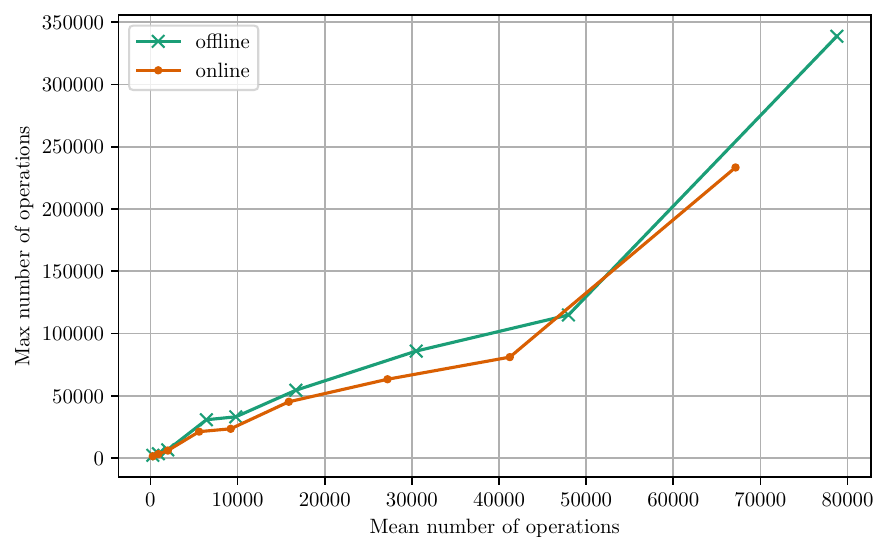}
        \caption{2D toric code with $L=(7,9,\dots,23)$.}
    \end{subfigure}
    \begin{subfigure}{\linewidth}
        \includegraphics[width=\linewidth]{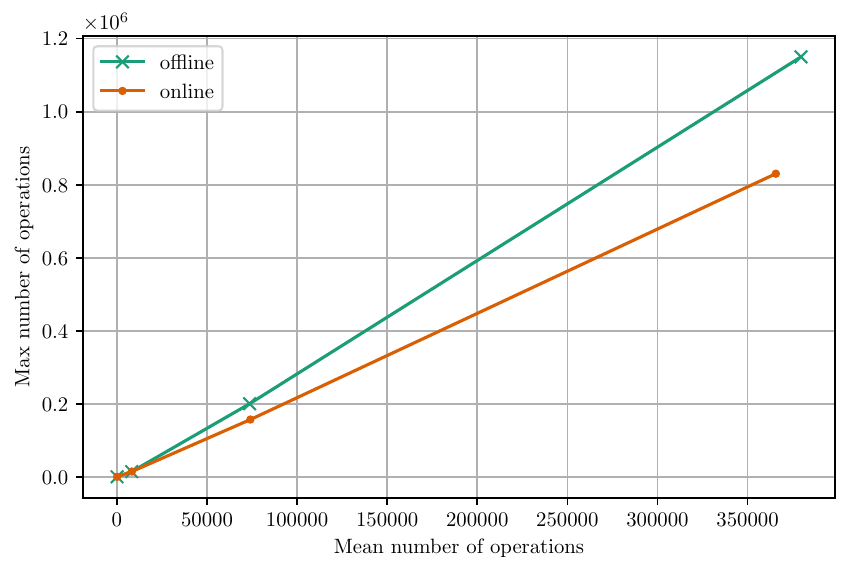}
        \caption{3D toric code with $L=(3,5,\dots,11)$.}
    \end{subfigure}
    \begin{subfigure}{\linewidth}
        \includegraphics[width=\linewidth]{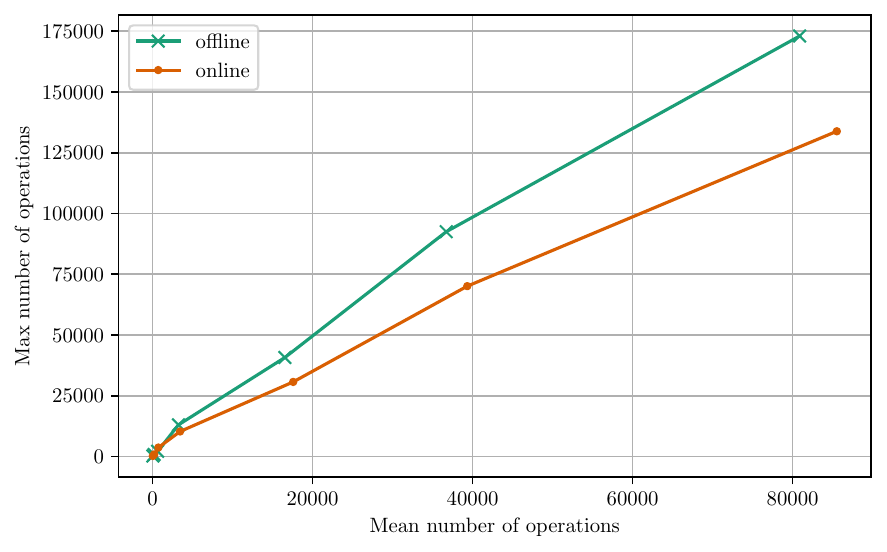}
        \caption{2D colour code with $n=(18, 36, 72, 144, 288, 432, 648)$.}
    \end{subfigure}
    
    \caption{Parametric plots showing the maximum and mean number of operations performed by offline and online decoders for increasing system size. Data for three different codes are shown: the 2D toric code (a), the 3D toric code (b), and the 2D $6.6.6$ colour code (c), all generated using $p=0.05$ and 60 shots per point.}
    \label{fig:operations}
\end{figure}

\begin{figure}
    \centering

    \begin{subfigure}{\linewidth}
        \includegraphics[width=\linewidth]{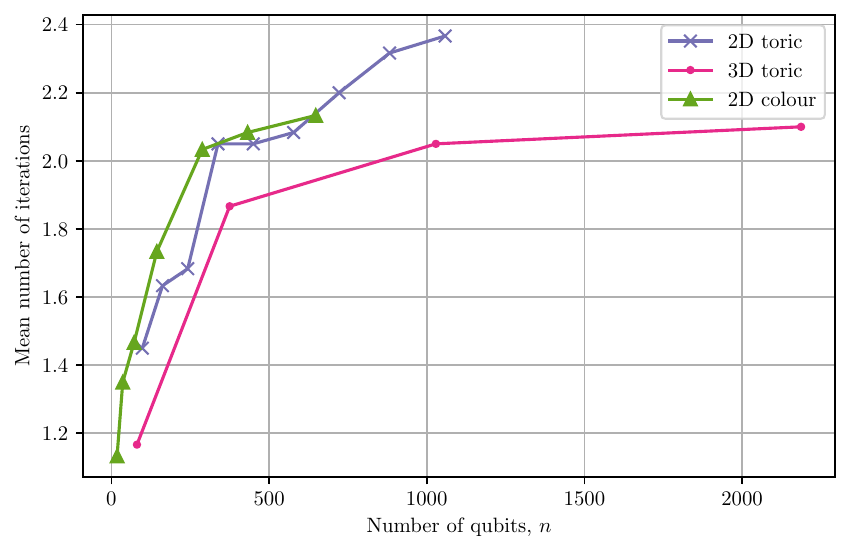}
        \caption{}
    \end{subfigure}
    \begin{subfigure}{\linewidth}
        \includegraphics[width=\linewidth]{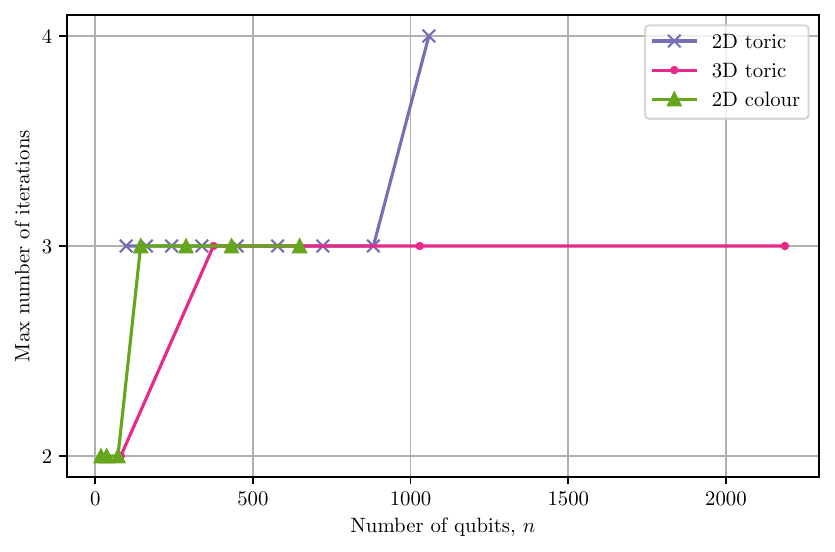}
        \caption{}
    \end{subfigure}
    
    \caption{Plots showing the mean (a) and maximum (b) number of iterations performed by the same decoding instances as in Fig.~\ref{fig:operations}. The numbers of qubits for the 2D and 3D toric codes are obtained via $2L^2$ and $3L^3$, respectively.}
    \label{fig:iterations}
\end{figure}

\begin{figure}
    \centering
    \includegraphics[width=\linewidth]{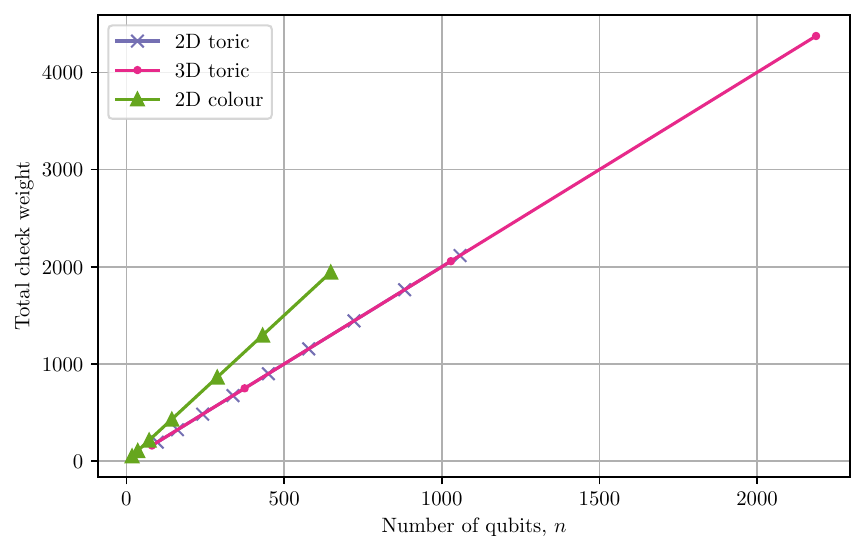}
    \caption{Total parity-check matrix weight for the three codes studied above.}
    \label{fig:check-weight}
\end{figure}

The complexity analysis above appears to show a well-defined speed-up for the online decoder versus the offline decoder. However, this relies on certain assumptions as to the exact behaviour of Gaussian elimination on the decoding instances. To illustrate this, we performed Monte Carlo simulations on three different code constructions: the 2D toric code, 3D toric code and 2D $6.6.6$ colour code with periodic boundary conditions. Note that it was shown in \cite{delfosse2022toward} that the toric codes are well-behaved as per the definition above, but this does not trivially extend to the colour code despite the structural similarity of these code families.

Fig.~\ref{fig:operations} shows the mean and maximum number of XOR operations performed by both online and offline decoders on code instances of increasing size. Specifically, we demonstrate decoding the $X$ stabilizer measurements under independent noise with $p=0.05$ with 60 shots per data point. Firstly, it is clear that across all three codes, the max number of operations is reduced by the online variant for increasing $n$, i.e.\ the worst-case complexity is invariably improved. This is broadly consistent with the complexity analysis above, which suggests a polynomially-scaling reduction in the number of operations. Secondly, the behaviour of the mean number of operations, i.e.\ the average-case complexity, is more nuanced. The mean number of operations also clearly improves for the 2D toric code, although this improvement is more slight for the 3D toric code -- meanwhile, the online decoder actually performs worse on this metric for the 2D colour code. To understand why, we empirically determine the covering radius for these same instances by recording the number of growth iterations, as shown in Fig.~\ref{fig:iterations}. By the nature of the online update, it is intuitive that a greater improvement should correlate with a higher number of iterations; that is, we expect the improvement to be starker on codes with a higher covering radius. The 2D toric code demonstrates the highest covering radius for increasing $n$, so it comes as no surprise that it should see the greatest improvement in average-case complexity. Whilst the complexity analysis suggests that the online variant should perform strictly faster regardless of covering radius, the reality is that retaining information between growth iterations may lead to suboptimal choices in pivot selection. The Gaussian elimination procedure has fewer rows (and thus pivots) to choose from at each stage on the smaller systems of earlier cluster growth cycles and -- unlike in the original offline implementation -- the decisions made by it are permanent. This can subtly increase the number of operations required at later growth cycles when filling missing pivots; the effect is dwarfed by the asymptotic improvement for higher covering radii, but can be significant for lower covering radii.

Finally, however, we see that the mean covering radius for the 2D colour code begins to converge somewhere between that of the other two codes, and yet it displays the worst performance for the online decoder. This can be attributed to the fact that, despite the number of iterations averaging between those of the toric codes, the mean number of operations is higher compared to the toric codes for similar values of $n$. This is demonstrated more clearly in Fig.~\ref{fig:check-weight}, which shows the total weight of the parity-check matrices (i.e.\ the total number of ones) for each code with increasing $n$. The 2D colour code outpaces both toric codes, demanding a greater number of operations for a similar covering radius; we can thus expect any suboptimal choices made by the online decoder to have a greater impact on its performance. This, however, is promising, as it suggests that this technique performs better for sparser parity-check matrices, which more closely represent the LDPC codes it is intended for.

\section{Conclusion} \label{sec:conclusion}
In this work, we have introduced an online variant of the Gaussian elimination subroutine used in qLDPC decoding. We have shown that -- in theory -- one can expect a polynomially-increasing reduction in the number of operations required compared to a standard offline implementation via complexity analysis inspired by the framework in \cite{delfosse2022toward}. While this asymptotic improvement competes with the negative effects of making suboptimal choices in pivot selection, we have shown that our online variant still outperforms offline in codes with a sparser parity-check matrix (i.e.\ LDPC codes) and higher covering radius.

During the completion of this article, the authors became aware of recently published work by Hillmann et al.\ in \cite{hillmannLocalizedStatisticsDecoding2024} where a related technique is discussed. 

\section*{Acknowledgements}
SJG was supported by University College London, Riverlane and the Engineering and Physical Sciences Research Council [grant number EP/S021582/1]. AB was supported by Engineering and Physical Sciences Research Council [grant numbers EP/Y004310/1 and EP/T001062/1]. DEB was supported by  Engineering and Physical Sciences Research Council [grant numbers EP/Y004310/1,  EP/T001062/1 and EP/Y004620/1].

\begin{figure*}
\begin{minipage}{\linewidth}
\begin{algorithm}[H]
    \caption{LUP decomposition in $\mathbb{Z}_2$}
    \label{alg:lup}
    \begin{algorithmic}[1]
        \Require $m \times n$ matrix $A$
        \Ensure Matrices $L,U,P$ satisfying $PA=LU$, where $L$ and $P$ are $m \times m$, and $U$ is $m \times n$ in row echelon form
        \State $L \gets 0_m$
        \State $U \gets A$
        \State $P \gets I_m$
        \State $r \gets 0, \ c \gets 0$
        \While{$r < m \AND c < n$}
            \State $i \gets r$ \Comment{Find next row with 1 in column $c$}
            \While{$i < m \AND U[i,c] = 0$}
                \State $i \gets i+1$
            \EndWhile

            \If{$i = m$} \Comment{If column had only zeroes left, move on to next column}
                \State $c \gets c+1$
            \Else
                \State Swap $L\text{.row}[i]$ and $L\text{.row}[r]$
                \State Swap $U\text{.row}[i]$ and $U\text{.row}[r]$
                \State Swap $P\text{.row}[i]$ and $P\text{.row}[r]$

                \State $i \gets r+1$ \Comment{Set ones beneath in this column by XORing rows}
                \While{$i < m$}
                    \If{$U[i,c]=1$}
                        \State $U\text{.row}[i] \gets U\text{.row}[i] \oplus U\text{.row}[r]$
                        \State $L[i,r] \gets 1$
                    \EndIf
                    \State $i \gets i+1$
                \EndWhile

                \State $r \gets r+1, \ c \gets c+1$ \Comment{Move on to next row and column}
            \EndIf
        \EndWhile

        \State $L \gets L + I_m$ \Comment{Set leading diagonal to ones}\footnote{$PA=LU$ is satisfied when $L$ is unit-triangular, i.e.\ has ones along its leading diagonal. For online update, this would require repeatedly subtracting/adding $I$ at the start/end of every iteration. This is merely a mathematical constraint rather than storing meaningful information, therefore it is most efficient to skip this entirely for the online variant.}
    \end{algorithmic}
\end{algorithm}
\end{minipage}
\end{figure*}

\begin{figure*}
\begin{minipage}{\linewidth}
\begin{algorithm}[H]
    \caption{Online LUP decomposition update in $\mathbb{Z}_2$}
    \label{alg:online-lup}
    \begin{algorithmic}[1]
        \Require Matrices $L,U,P$ where all have $r_\text{new}$ new rows and $U$ has $c_\text{new}$ new columns
        \Ensure Updated matrices $L,U,P$ satisfying $PA=LU$, with $U$ returned to row echelon form
        
        \State $U[\text{old rows, new cols}] \gets PU[\text{old rows, new cols}]$ \Comment{Swap new cols in old rows from old $P$}

        \State $r \gets 1$ \Comment{XOR new cols according to $L$}
        \While{$r < r_\text{old}$}
            \State $c \gets 0$
            \While{$c < r$}
                \If{$L[r,c]=1$}
                    \State $U[r, \text{new cols}] \gets U[r, \text{new cols}] \oplus U[c, \text{new cols}]$
                \EndIf
                \State $c \gets c+1$
            \EndWhile
            \State $r \gets r+1$
        \EndWhile

        \State $r,c \gets 0$ \Comment{Recommence Gaussian elimination procedure}
        \While{$r < m \AND c < n$}
            \State $i \gets r$ \Comment{Find next row with 1 in column $c$}
            \If{$U[i,c]=0$} \Comment{(skip to new rows if still in old columns)}
                \If{$c<c_\text{old}$}
                    \State $i \gets \max(r_\text{old}, r+1)$
                \Else
                    \State $i \gets r+1$
                \EndIf
                \While{$i < m \AND U[i,c]=0$}
                    \State $i \gets i+1$
                \EndWhile
            \EndIf
            \If{$i=m$} \Comment{If column had only zeroes left, move on to next column}
                \State $c \gets c+1$
            \Else
                \State Swap $L\text{.row}[i]$ and $L\text{.row}[r]$
                \State Swap $U\text{.row}[i]$ and $U\text{.row}[r]$
                \State Swap $P\text{.row}[i]$ and $P\text{.row}[r]$

                \If{$c<c_\text{old}$} \Comment{Set ones beneath in this column by XORing rows}
                    \State $i \gets \max(r_\text{old}, r+1)$ \Comment{(skip to new rows if still in old columns)}
                \Else
                    \State $i \gets r+1$
                \EndIf
                \While{$i < m$}
                    \If{$U[i,c]=1$}
                        \State $U\text{.row}[i] \gets U\text{.row}[i] \oplus U\text{.row}[r]$
                        \State $L[i,r] \gets 1$
                    \EndIf
                    \State $i \gets i+1$
                \EndWhile

                \State $r \gets r+1, \ c \gets c+1$ \Comment{Move on to next row and column}
            \EndIf
        \EndWhile
    \end{algorithmic}
\end{algorithm}
\end{minipage}
\end{figure*}

\newpage
\printbibliography

@book{shafarevichLinearAlgebraGeometry2012,
  title = {Linear Algebra and Geometry},
  author = {Shafarevich, Igor R. and Remizov, Alexey O.},
  date = {2012-08-23},
  publisher = {Springer Science \& Business Media},
  isbn = {978-3-642-30994-6},
  langid = {english},
  pagetotal = {536}
}

@inproceedings{shor1996fault,
  title = {Fault-Tolerant Quantum Computation},
  booktitle = {Proceedings of 37th {{Conference}} on {{Foundations}} of {{Computer Science}}},
  author = {Shor, P.W.},
  date = {1996-10},
  pages = {56--65},
  issn = {0272-5428},
  doi = {10.1109/SFCS.1996.548464},
  url = {https://ieeexplore.ieee.org/abstract/document/548464},
  urldate = {2024-04-22},
  eventtitle = {Proceedings of 37th {{Conference}} on {{Foundations}} of {{Computer Science}}}
}

@incollection{preskill1998fault,
  title = {Fault-Tolerant Quantum Computation},
  booktitle = {Introduction to {{Quantum Computation}} and {{Information}}},
  author = {Preskill, John},
  date = {1998-10},
  pages = {213--269},
  publisher = {World Scientific},
  doi = {10.1142/9789812385253_0008},
  url = {https://www.worldscientific.com/doi/abs/10.1142/9789812385253_0008},
  urldate = {2024-04-22},
  isbn = {978-981-02-3399-0}
}

@article{dennis2002topological,
  title = {Topological Quantum Memory},
  author = {Dennis, Eric and Kitaev, Alexei and Landahl, Andrew and Preskill, John},
  date = {2002-09},
  journaltitle = {Journal of Mathematical Physics},
  shortjournal = {Journal of Mathematical Physics},
  volume = {43},
  number = {9},
  eprint = {quant-ph/0110143},
  eprinttype = {arXiv},
  pages = {4452--4505},
  issn = {0022-2488, 1089-7658},
  doi = {10.1063/1.1499754},
  url = {http://arxiv.org/abs/quant-ph/0110143},
  urldate = {2020-05-30}
}

@article{ai2024quantum,
  title = {Quantum Error Correction below the Surface Code Threshold},
  author = {{Google Quantum AI and Collaborators}},
  date = {2025},
  journaltitle = {Nature},
  shortjournal = {Nature},
  volume = {638},
  number = {8052},
  eprint = {39653125},
  eprinttype = {pmid},
  pages = {920--926},
  issn = {0028-0836},
  doi = {10.1038/s41586-024-08449-y},
  url = {https://www.ncbi.nlm.nih.gov/pmc/articles/PMC11864966/},
  urldate = {2025-04-03},
  pmcid = {PMC11864966}
}

@article{delfosse2021almost,
  title = {Almost-Linear Time Decoding Algorithm for Topological Codes},
  author = {Delfosse, Nicolas and Nickerson, Naomi H.},
  year = {2021},
  month = dec,
  journal = {Quantum},
  volume = {5},
  pages = {595},
  publisher = {Verein zur F{\"o}rderung des Open Access Publizierens in den Quantenwissenschaften},
  doi = {10.22331/q-2021-12-02-595},
  urldate = {2025-03-16},
  langid = {british}
}

@article{tillich2013quantum,
  title = {Quantum {{LDPC}} Codes with Positive Rate and Minimum Distance Proportional to the Square Root of the Blocklength},
  author = {Tillich, Jean-Pierre and Zémor, Gilles},
  date = {2014-02},
  journaltitle = {IEEE Transactions on Information Theory},
  volume = {60},
  number = {2},
  pages = {1193--1202},
  issn = {1557-9654},
  doi = {10.1109/TIT.2013.2292061},
  url = {https://ieeexplore.ieee.org/document/6671468},
  urldate = {2025-03-16},
  eventtitle = {{{IEEE Transactions}} on {{Information Theory}}}
}

@article{panteleevDegenerateQuantumLDPC2021,
  title = {Degenerate Quantum {{LDPC}} Codes with Good Finite Length Performance},
  author = {Panteleev, Pavel and Kalachev, Gleb},
  date = {2021-11-22},
  journaltitle = {Quantum},
  volume = {5},
  pages = {585},
  publisher = {Verein zur Förderung des Open Access Publizierens in den Quantenwissenschaften},
  doi = {10.22331/q-2021-11-22-585},
  url = {https://quantum-journal.org/papers/q-2021-11-22-585/},
  urldate = {2024-04-18},
  langid = {british}
}

@inproceedings{panteleev2022asymptotically,
  title = {Asymptotically Good Quantum and Locally Testable Classical {{LDPC}} Codes},
  booktitle = {Proceedings of the 54th {{Annual ACM SIGACT Symposium}} on {{Theory}} of {{Computing}}},
  author = {Panteleev, Pavel and Kalachev, Gleb},
  date = {2022-06-10},
  series = {{{STOC}} 2022},
  pages = {375--388},
  publisher = {Association for Computing Machinery},
  location = {New York, NY, USA},
  doi = {10.1145/3519935.3520017},
  url = {https://doi.org/10.1145/3519935.3520017},
  urldate = {2025-03-12},
  isbn = {978-1-4503-9264-8}
}

@article{mceliece1998turbo,
  title = {Turbo Decoding as an Instance of {{Pearl}}'s ``Belief Propagation'' Algorithm},
  author = {McEliece, R.J. and MacKay, D.J.C. and Cheng, Jung-Fu},
  date = {1998-02},
  journaltitle = {IEEE Journal on Selected Areas in Communications},
  volume = {16},
  number = {2},
  pages = {140--152},
  issn = {1558-0008},
  doi = {10.1109/49.661103},
  url = {https://ieeexplore.ieee.org/abstract/document/661103},
  urldate = {2024-04-22},
  eventtitle = {{{IEEE Journal}} on {{Selected Areas}} in {{Communications}}}
}

@article{mackay1997near,
  title = {Near {{Shannon}} Limit Performance of Low Density Parity Check Codes},
  author = {MacKay, D.J.C. and Neal, R.M.},
  date = {1996-08-29},
  journaltitle = {Electronics Letters},
  volume = {32},
  number = {18},
  pages = {1645--1646},
  publisher = {{The Institution of Engineering and Technology}},
  doi = {10.1049/el:19961141},
  url = {https://digital-library.theiet.org/doi/10.1049/el%3A19961141},
  urldate = {2025-03-16}
}

@article{richardson2001capacity,
  title = {The Capacity of Low-Density Parity-Check Codes under Message-Passing Decoding},
  author = {Richardson, T.J. and Urbanke, R.L.},
  date = {2001-02},
  journaltitle = {IEEE Transactions on Information Theory},
  volume = {47},
  number = {2},
  pages = {599--618},
  issn = {1557-9654},
  doi = {10.1109/18.910577},
  url = {https://ieeexplore.ieee.org/document/910577},
  urldate = {2025-03-16},
  eventtitle = {{{IEEE Transactions}} on {{Information Theory}}}
}

@article{roffeDecodingQuantumLDPC2020,
  title = {Decoding across the Quantum {{LDPC}} Code Landscape},
  author = {Roffe, Joschka and White, David R. and Burton, Simon and Campbell, Earl T.},
  date = {2020-12-28},
  journaltitle = {Physical Review Research},
  shortjournal = {Phys. Rev. Research},
  volume = {2},
  number = {4},
  eprint = {2005.07016},
  eprinttype = {arXiv},
  eprintclass = {quant-ph},
  pages = {043423},
  issn = {2643-1564},
  doi = {10.1103/PhysRevResearch.2.043423},
  url = {http://arxiv.org/abs/2005.07016},
  urldate = {2024-02-23}
}

@book{macwilliamsTheoryErrorcorrectingCodes1977,
  title = {The Theory of Error-Correcting Codes},
  author = {MacWilliams, Florence Jessie and Sloane, Neil James Alexander},
  date = {1977},
  eprint = {nv6WCJgcjxcC},
  eprinttype = {googlebooks},
  publisher = {Elsevier},
  isbn = {978-0-444-85010-2},
  langid = {english},
  pagetotal = {788}
}

@article{delfosse2022toward,
  title = {Toward a Union-Find Decoder for Quantum {{LDPC}} Codes},
  author = {Delfosse, Nicolas and Londe, Vivien and Beverland, Michael E.},
  date = {2022-05},
  journaltitle = {IEEE Transactions on Information Theory},
  volume = {68},
  number = {5},
  pages = {3187--3199},
  issn = {1557-9654},
  doi = {10.1109/TIT.2022.3143452},
  eventtitle = {{{IEEE Transactions}} on {{Information Theory}}}
}

@article{fossorierSoftDecisionDecoding1997,
  title = {Soft Decision Decoding of Linear Block Codes Based on Ordered Statistics for the {{Rayleigh}} Fading Channel with Coherent Detection},
  author = {Fossorier, M.P.C. and Lin, Shu},
  date = {1997-01},
  journaltitle = {IEEE Transactions on Communications},
  volume = {45},
  number = {1},
  pages = {12--14},
  issn = {1558-0857},
  doi = {10.1109/26.554278},
  url = {https://ieeexplore.ieee.org/document/554278},
  urldate = {2024-04-18},
  eventtitle = {{{IEEE Transactions}} on {{Communications}}}
}

@article{delfosse2020linear,
  title = {Linear-Time Maximum Likelihood Decoding of Surface Codes over the Quantum Erasure Channel},
  author = {Delfosse, Nicolas and Zémor, Gilles},
  date = {2020-07-09},
  journaltitle = {Physical Review Research},
  shortjournal = {Phys. Rev. Res.},
  volume = {2},
  number = {3},
  pages = {033042},
  publisher = {American Physical Society},
  doi = {10.1103/PhysRevResearch.2.033042},
  url = {https://link.aps.org/doi/10.1103/PhysRevResearch.2.033042},
  urldate = {2024-04-12}
}

@article{griffithsUnionfindQuantumDecoding2024,
  title = {Union-Find Quantum Decoding without Union-Find},
  author = {Griffiths, Sam J. and Browne, Dan E.},
  date = {2024-02-09},
  journaltitle = {Physical Review Research},
  shortjournal = {Phys. Rev. Res.},
  volume = {6},
  number = {1},
  pages = {013154},
  publisher = {arXiv: 2306.09767},
  doi = {10.1103/PhysRevResearch.6.013154},
  url = {https://link.aps.org/doi/10.1103/PhysRevResearch.6.013154},
  urldate = {2024-03-07}
}

@book{borodinOnlineComputationCompetitive2005,
  title = {Online Computation and Competitive Analysis},
  author = {Borodin, Allan and El-Yaniv, Ran},
  date = {2005-02-17},
  eprint = {v3faI8pER6IC},
  eprinttype = {googlebooks},
  publisher = {Cambridge University Press},
  isbn = {978-0-521-61946-2},
  langid = {english},
  pagetotal = {440}
}

@book{boydIntroductionAppliedLinear2018,
  title = {Introduction to Applied Linear Algebra: Vectors, Matrices, and Least Squares},
  shorttitle = {Introduction to Applied Linear Algebra},
  author = {Boyd, Stephen and Vandenberghe, Lieven},
  date = {2018-06-07},
  eprint = {IApaDwAAQBAJ},
  eprinttype = {googlebooks},
  publisher = {Cambridge University Press},
  isbn = {978-1-316-51896-0},
  langid = {english},
  pagetotal = {477}
}

@online{hillmannLocalizedStatisticsDecoding2024,
  title = {Localized Statistics Decoding: {{A}} Parallel Decoding Algorithm for Quantum Low-Density Parity-Check Codes},
  shorttitle = {Localized Statistics Decoding},
  author = {Hillmann, Timo and Berent, Lucas and Quintavalle, Armanda O. and Eisert, Jens and Wille, Robert and Roffe, Joschka},
  date = {2024-06-26},
  eprint = {2406.18655},
  eprinttype = {arXiv},
  eprintclass = {quant-ph},
  doi = {10.48550/arXiv.2406.18655},
  url = {http://arxiv.org/abs/2406.18655},
  pubstate = {prepublished}
}

@book{nielsenQuantumComputationQuantum2002,
  title = {Quantum Computation and Quantum Information},
  author = {Nielsen, Michael A. and Chuang, Isaac L.},
  date = {2002-04-12},
  publisher = {Cambridge University Press},
  url = {https://aapt.scitation.org/doi/10.1119/1.1463744},
  urldate = {2019-01-16}
}

@inproceedings{gottesmanIntroductionQuantumError2002,
  title = {An Introduction to Quantum Error Correction},
  booktitle = {Proceedings of {{Symposia}} in {{Applied Mathematics}}},
  author = {Gottesman, Daniel},
  date = {2002},
  volume = {58},
  eprint = {0904.2557},
  eprinttype = {arXiv},
  eprintclass = {quant-ph},
  pages = {221--236}
}

@article{kitaevFaulttolerantQuantumComputation2003,
  title = {Fault-Tolerant Quantum Computation by Anyons},
  author = {Kitaev, A. Yu},
  date = {2003-01},
  journaltitle = {Annals of Physics},
  shortjournal = {Annals of Physics},
  volume = {303},
  number = {1},
  eprint = {quant-ph/9707021},
  eprinttype = {arXiv},
  pages = {2--30},
  issn = {00034916},
  doi = {10.1016/S0003-4916(02)00018-0},
  url = {http://arxiv.org/abs/quant-ph/9707021},
  urldate = {2025-03-18}
}

@article{castelnovo2008topological,
  title = {Topological Order in a Three-Dimensional Toric Code at Finite Temperature},
  author = {Castelnovo, Claudio and Chamon, Claudio},
  date = {2008-10-21},
  journaltitle = {Physical Review B},
  shortjournal = {Phys. Rev. B},
  volume = {78},
  number = {15},
  pages = {155120},
  publisher = {American Physical Society},
  doi = {10.1103/PhysRevB.78.155120},
  url = {https://link.aps.org/doi/10.1103/PhysRevB.78.155120},
  urldate = {2025-03-31}
}

@article{bombin2006topological,
  title = {Topological Quantum Distillation},
  author = {Bombin, H. and Martin-Delgado, M. A.},
  date = {2006-10-30},
  journaltitle = {Physical Review Letters},
  shortjournal = {Phys. Rev. Lett.},
  volume = {97},
  number = {18},
  pages = {180501},
  publisher = {American Physical Society},
  doi = {10.1103/PhysRevLett.97.180501},
  url = {https://link.aps.org/doi/10.1103/PhysRevLett.97.180501},
  urldate = {2025-04-03}
}

@article{skoric2023parallel,
  title = {Parallel Window Decoding Enables Scalable Fault Tolerant Quantum Computation},
  author = {Skoric, Luka and Browne, Dan E. and Barnes, Kenton M. and Gillespie, Neil I. and Campbell, Earl T.},
  date = {2023-11-03},
  journaltitle = {Nature Communications},
  shortjournal = {Nat Commun},
  volume = {14},
  number = {1},
  pages = {7040},
  publisher = {Nature Publishing Group},
  issn = {2041-1723},
  doi = {10.1038/s41467-023-42482-1},
  url = {https://www.nature.com/articles/s41467-023-42482-1},
  urldate = {2024-01-12},
  issue = {1},
  langid = {english}
}

@article{calderbank1996good,
  title = {Good Quantum Error-Correcting Codes Exist},
  author = {Calderbank, A. R. and Shor, Peter W.},
  date = {1996-08-01},
  journaltitle = {Physical Review A},
  shortjournal = {Phys. Rev. A},
  volume = {54},
  number = {2},
  eprint = {quant-ph/9512032},
  eprinttype = {arXiv},
  pages = {1098--1105},
  issn = {1050-2947, 1094-1622},
  doi = {10.1103/PhysRevA.54.1098},
  url = {http://arxiv.org/abs/quant-ph/9512032},
  urldate = {2022-02-05}
}

@article{terhal2015quantum,
  title = {Quantum Error Correction for Quantum Memories},
  author = {Terhal, Barbara M.},
  date = {2015},
  journaltitle = {Reviews of Modern Physics},
  volume = {87},
  number = {2},
  pages = {307--346},
  doi = {10.1103/RevModPhys.87.307}
}

\end{document}